\documentclass[aps,prb,twocolumn,superscriptaddress,amsmath,amssymb]{revtex4}

\usepackage{amssymb}
\usepackage[dvips]{graphicx}
\usepackage{bm}
\usepackage{epsfig}
\usepackage[ansinew]{inputenc}

\begin{document}



\title{Compensation temperatures and exchange bias in La$_{1.5}$Ca$_{0.5}$CoIrO$_{6}$}

\author{L. T. Coutrim}
\affiliation{Instituto de F\'{\i}sica, Universidade Federal de Goi\'{a}s, 74001-970, Goi\^{a}nia, GO, Brazil}

\author{E. M. Bittar}
\affiliation{Centro Brasileiro de Pesquisas F\'{\i}sicas, 22290-180, Rio de Janeiro, RJ, Brazil}

\author{F. Stavale}
\affiliation{Centro Brasileiro de Pesquisas F\'{\i}sicas, 22290-180, Rio de Janeiro, RJ, Brazil}

\author{F. Garcia}
\affiliation{Centro Brasileiro de Pesquisas F\'{\i}sicas, 22290-180, Rio de Janeiro, RJ, Brazil}

\author{E. Baggio-Saitovitch}
\affiliation{Centro Brasileiro de Pesquisas F\'{\i}sicas, 22290-180, Rio de Janeiro, RJ, Brazil}

\author{M. Abbate}
\affiliation{Universidade Federal do Paran\'{a}, 19044, 81531-990 Curitiba, PR, Brazil}

\author{R. J. O. Mossanek}
\affiliation{Universidade Federal do Paran\'{a}, 19044, 81531-990 Curitiba, PR, Brazil}

\author{H. P. Martins}
\affiliation{Universidade Federal do Paran\'{a}, 19044, 81531-990 Curitiba, PR, Brazil}

\author{D. Tobia}
\affiliation{Instituto de F\'{\i}sica ``Gleb Wataghin", UNICAMP, 13083-859, Campinas, SP, Brazil}

\author{P. G. Pagliuso}
\affiliation{Instituto de F\'{\i}sica ``Gleb Wataghin", UNICAMP, 13083-859, Campinas, SP, Brazil}

\author{L. Bufai\c{c}al}
\email{lbufaical@ufg.br}
\affiliation{Instituto de F\'{\i}sica, Universidade Federal de Goi\'{a}s, 74001-970, Goi\^{a}nia, GO, Brazil}

\date{\today}

\begin{abstract}
We report on the study of magnetic properties of the La$_{1.5}$Ca$_{0.5}$CoIrO$_{6}$ double perovskite. Via ac magnetic susceptibility we have observed
evidence of weak ferromagnetism and reentrant spin glass behavior on an antiferromagnetic matrix. Regarding the magnetic behavior as a function of
temperature, we have found that the material displays up to three inversions of its magnetization, depending on the appropriate choice of the applied
magnetic field. At low temperature the material exhibit exchange bias effect when it is cooled in the presence of a magnetic field. Also, our results
indicate that this effect may be observed even when the system is cooled at zero field. Supported by other measurements and also by electronic structure
calculations, we discuss the magnetic reversals and spontaneous exchange bias effect in terms of magnetic phase separation and magnetic frustration of
Ir$^{4+}$ ions located between the antiferromagnetically coupled Co ions.
\end{abstract}


\maketitle

\section{Introduction}

Magnetic frustration emerges from competing magnetic interactions and degenerate multivalley ground states \cite{Knolle}. An example of such systems are
the spin glasses (SG), in which the interactions between magnetic moments are in conflict with each other due to the presence of frozen-in structural
disorder. The intriguing physical phenomena underlying the SG behavior have led to great interest in these materials since the 1970's. The development of
theories to model the SG found its applicability in a wide variety of research fields, from real glasses to neural networks and protein folding
\cite{Binder,Mydosh}. Despite the great scientific interest and intense research over the last decades, the underlying physics that govern the SG phenomena
is far from being well understood.

The canonical example of a SG type material is an intermetallic alloy in which a few percent of magnetic ions are dispersed arbitrarily in a non-magnetic
matrix. These magnetic atoms are therefore separated by incidental distances, and thus the RKKY (Ruderman-Kittel-Kasuya-Yosida) interaction allows the
coupling energy to have random sign. This class of systems corresponds to the historical discovery of SG, which traces back to the studies of strongly
diluted magnetic alloys and the Kondo effect \cite{Mydosh}.

Later on, SG have been identified within other systems, such as insulating intermetallics, layered thin films and magnetic oxides. For the later class of
materials, is important to mention the geometric frustrated pyrochlores\cite{Lee,Greedan} and the extensively studied Co/Mn-based perovskites
\cite{Teresa,Kundu,Xu}. For these A(Co,Mn)O$_{3}$ materials, the SG-like behavior is usually ascribed to the phase segregation and inhomogeneity of the
compounds, and the dynamical process of magnetic relaxation being related to the growth and interactions of magnetic clusters \cite{Binder,Salamon}. Many
of these cobaltites and manganites exhibit the interesting exchange-bias (EB) effect, for which there is a shift of the magnetization as a function of the
applied magnetic field [$M(H)$] curve in respect to its center. This phenomena is usually attributed to the induced exchange anisotropy at the interface
between antiferromagnetic (AFM) and ferromagnetic (FM)/ferrimagnetic (FIM) phases in heterogeneous systems \cite{Nogues}, and it has also been observed at
FM/FIM/AFM-SG interfaces \cite{Mannan,Gruyters}. It is commonly observed in multilayered systems, although also found in bulk materials with competing
magnetic interactions with undefined magnetic interfaces between regions with AFM or FIM interactions \cite{Nogues}.

We recently reported the structural and magnetic characterization of the La$_{2-x}$Ca$_{x}$CoIrO$_{6}$ double perovskite series \cite{JSSC}. For
La$_{2}$CoIrO$_{6}$ the transition-metal ions valences are reported to be Co$^{2+}$ and Ir$^{4+}$ \cite{Kolchinskaya}. Our results show that La$^{3+}$ to
Ca$^{2+}$ substitution leads to Co valence changes, and La$_{1.5}$Ca$_{0.5}$CoIrO$_{6}$ presents both Co$^{2+}$ and Co$^{3+}$ in high spin configuration.
Moreover, it presents the two key ingredients to achieve a SG state, which are different competing magnetic interactions and disorder
\cite{Binder,Mydosh,JSSC}.

In the course of studying the magnetic properties of La$_{1.5}$Ca$_{0.5}$CoIrO$_{6}$, we have observed a reentrant spin glass (RSG)-like state,
\textit{i.e.}, there is a conventional magnetic ordering and, at lower temperature ($T$), the system achieves the SG state concomitantly to other magnetic
phases. This behavior is ascribed to the Ir magnetic frustration due to Co AFM coupling and to the magnetic phase separation induced by the anti-site
disorder (ASD) at the transition-metal sites.

This RSG state is intrinsically related to the EB effect observed for La$_{1.5}$Ca$_{0.5}$CoIrO$_{6}$. Another important feature observed is that, for an
appropriate choice of the applied dc magnetic field ($H_{dc}$) the compound can reverse its magnetization up to three times in the $T$-dependent
magnetization measurements.

In this work we report the investigation of the magnetic behavior of La$_{1.5}$Ca$_{0.5}$CoIrO$_{6}$. Using ac magnetic susceptibility measurements we have
observed that the system exhibit conventional magnetic orderings at $T\sim90$ K and a spin glass transition at $T\simeq27$ K, confirming its RSG state at
low-$T$. The dc magnetization measurements revealed that the high-$T$ anomaly is in fact associated with two conventional magnetic transitions, the AFM and
FM phases of Co ions. The compensation temperatures in the $T$-dependent magnetization measurements and the spontaneous EB effect in the isothermal
magnetization measurements were investigated in detail. Studies of x-ray absorption near edge structure (XANES), x-ray magnetic circular dichroism (XMCD),
x-ray photoelectron spectroscopy (XPS) and band structure calculations were also carried out. These data corroborates our argument that the magnetization
reversals and zero field cooled (ZFC) exchange bias effect observed for La$_{1.5}$Ca$_{0.5}$CoIrO$_{6}$ can be both understood in terms of the same
underlying mechanism, \textit{i.e.}, are due to the competing interactions of Co ions that leads to magnetic phase segregation and frustration of the Ir
magnetic moments.

\section{Experimental details}

Polycrystalline samples of La$_{1.5}$Ca$_{0.5}$CoIrO$_{6}$ were prepared by the solid state reaction in a conventional tubular furnace and air atmosphere.
Stoichiometric amounts of La$_{2}$O$_{3}$, CaO, Co$_{3}$O$_{4}$ and metallic Ir were mixed and heated at $650^{\circ}$C for 24 hours. Later the samples
were re-grinded before a second step of 48 hours at $800^{\circ}$C. Finally the materials were grinded, pressed into pellets and heated at $975^{\circ}$C
for two weeks. X-ray powder diffraction pattern revealed a single phase double perovskite structure with monoclinic $P2_1/n$ symmetry, with 9\% of ASD at
Co/Ir sites \cite{JSSC}. Magnetic data were collected on a commercial Physical Property Measurement System. AC magnetic susceptibility was measured with
driving field $H_{ac}=10$ Oe, at the frequency range of 10-10000 Hz. DC magnetization was measured at ZFC and field cooled (FC) modes. XPS experiments were
performed using an ultra-high vacuum (UHV) chamber equipped with a SPECS analyzer PHOIBOS 150. XMCD and XANES measurements were performed in the dispersive
x-ray absorption (DXAS) beam line at the Brazilian Synchrotron Light Laboratory (LNLS) \cite{dxas}. The edge step normalization of the data was performed
after a linear pre-edge subtraction and the regression of a quadratic polynomial beyond the edge, using the software ATHENA \cite{athena}. The band
structure calculations were performed using the WIEN2k software package \cite{Wien2k}.

\section{Results and discussion}

\subsection{X-ray Photoelectron Spectroscopy (XPS)}

\begin{figure}
\begin{center}
\includegraphics[width=0.5 \textwidth]{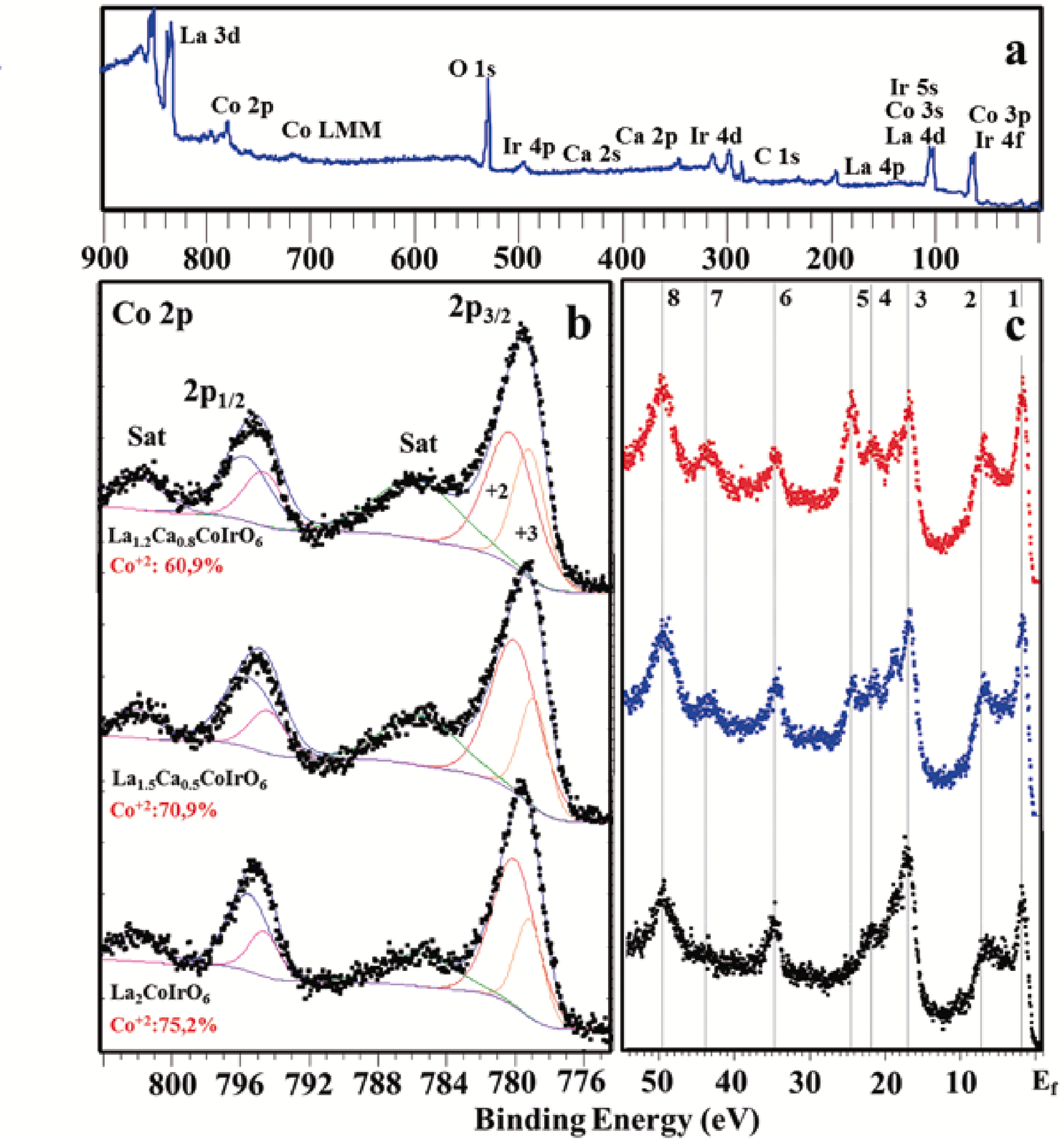}
\vspace{-0.8 cm}
\end{center}
\caption{(a) XPS survey spectra for La$_{1.5}$Ca$_{0.5}$CoIrO$_{6}$. All elements observed in the sample are indicated. (b) High resolution Co 2$p$ region
for several Ca doping levels and the corresponding peak components fitting using a Shirley background and two Lorentzian-Gaussian peaks for Co$^{2+}$ and
Co$^{3+}$ cations component. (c) The corresponding valence band photoemission features and the main peaks indicated by line 1 and line 2, related to the
high-spin state of the Co$^{3+}$ cations.}
\label{FigXPS}
\end{figure}

Fig. \ref{FigXPS} shows the XPS spectra for La$_{1.5}$Ca$_{0.5}$CoIrO$_{6}$. All the results presented correspond to the use of monochromatic Al
$K_{\alpha}$ x-ray radiation ($h\nu=1486.6$ eV) using spectrometer pass energy ($E_{pass}$) of 15 eV. The spectrometer was previously calibrated using the
Au 4$f_{7/2}$ (84.0 eV) which results on a full-width-half-maximum of 0.7 eV, for a sputtered metallic gold foil. The samples were referenced by setting
the adventitious carbon C 1$s$ peak to 284.6 eV. Prior to mounting in UHV, samples were slightly polished and ultra-sonicated sequentially in isopropyl
alcohol and water. The photo-emission spectra were sequentially acquired after successive cycles of gentle Ar$^{+}$ sputtering ($5\times10^{-7}$ mbar Ar,1
kV, 1 $\mu$A/cm$^2$, 10-15 seconds intervals) for hydrocarbons removal. All spectra shown correspond to 45 s Ar$^{+}$ sputtering at the ideally found
conditions to avoid surface species reduction.

In Fig. \ref{FigXPS}(a) all elements found are indicated. We observe that residual carbon persist on the samples surface since the in-situ sputtering has
been employed at the optimal conditions to avoid chemical reduction. Particularly important is the careful analysis of the cobalt 2$p$ region which serves
as a tool to assign Co cations sites and electronic configuration in several compounds \cite{1,2,3}. Noteworthy, although XPS is most related to the
surface chemical composition based on the limited photoelectrons inelastic mean free path in solids, because of the large mean free path of the cobalt
photoelectrons (for $E_{Kin}$$\sim700$ eV, about 1.5 nm) the spectra presented can certainly reveal additional information from buried layers. As reported
extensively in the literature, the expected binding energies for Co$^{3+}$ and Co$^{2+}$ are 779.5 and 780.5 eV, respectively \cite{4}. Importantly,
depending on the cation symmetry one may observe satellite features related to charge-transfer mechanisms which appear significantly stronger for Co$^{2+}$
in octahedral sites, in contrast to Co$^{3+}$ in octahedral or Co$^{2+}$ in tetrahedral sites. The differences has been discussed in the literature and
arises from an enhanced screening of Co$^{3+}$ due to a larger Co 3$d$-O 2$p$ hopping strength and a smaller charge-transfer energy compared to those in
the Co$^{2+}$ charge state \cite{5}. In Fig. \ref{FigXPS}, the Co 2$p$ XPS spectra show prominent peaks at 779.6 and 795.3 eV related to the Co 2$p_{3/2}$
and 2$p_{1/2}$, respectively, in addition to the satellite feature located at 785.3 eV. The intense satellite is a clear signature of Co$^{2+}$ in
octahedral sites, most probably at a high-spin Co$^{2+}$ configuration as observed for the cobalt monoxide \cite{6}.

The detailed analysis of the Co 2$p$ region is however far from trivial, and the main peaks are known to be composed by multiplet splitting as described by
Gupta and Sen \cite{Gupta}, and extensively discussed in several recently studies \cite{Ivanova,Biesinger}. Nevertheless, we have considered that the
intensity of each spin-orbit component can be fitted using two curves related to the different Co cations and respective satellites since Co$^{2+}$ and
Co$^{3+}$ are known to display well distinguish binding energies at about 780.1 and 779.6 eV \cite{Biesinger}. The spectral analysis for
La$_{1.5}$Ca$_{0.5}$CoIrO$_{6}$ compound employing two fitting components results on the assignment of Co$^{2+}$ and Co$^{3+}$ species located at binding
energies and full-width-half-maximum of 780.1 (3.03)  and 779.3 (1.89) eV, respectively. In order to compare the results obtained, the same procedure has
been employed for La$_{2}$CoIrO$_{6}$ and La$_{1.2}$Ca$_{0.8}$CoIrO$_{6}$, as depicted in Fig. \ref{FigXPS}(b). The results indicate a Co$^{2+}$/Co$^{3+}$
ratio of 2.33 for La$_{2}$CoIrO$_{6}$, 2.44 for La$_{1.5}$Ca$_{0.5}$CoIrO$_{6}$ and 3.03 for La$_{1.2}$Ca$_{0.8}$CoIrO$_{6}$. The increase of
Co$^{2+}$/Co$^{3+}$ ratio as a function of the Ca incorporation is expected and has served to test our fitting procedure. Certainly, the absolute amount
can be hardly determined. However, the trend observed can fairly indicate the relative amount of Co$^{3+}$ species in the compounds.

Furthermore, the photoemission spectra close to the valence band [Fig. \ref{FigXPS}(c)] is particularly helpful to distinguish the spin state of the Co
cations since it display noticeable multiplet structure. The spectral features observed have been discussed by several authors in great detail for CoO,
Co$_{3}$O$_{4}$ and more recently for LaCoO$_{3}$\cite{6,Raveau,tese}. The most important structure are those close to 1-8 eV range, indicated by line 1,
which is expected for Co$^{3+}$ in octahedral sites, and the one indicated by line 2, related to strong multiplet effects in the final state of Co$^{3+}$
cations. Previous calculations and photoemission studies have shown that Co$^{3+}$ in a purely, low-spin state is characterized by an intense peak close to
1 eV but broad features at energies ranging up to 8 eV \cite{Raveau}. In contrast, the La$_{1.5}$Ca$_{0.5}$CoIrO$_{6}$ compound displays two features
peaking at line 1 and line 2, which suggest a Co$^{3+}$ high spin state.

The approximate 70$\%$/30$\%$ proportion of Co$^{2+}$/Co$^{3+}$ estimated from the XPS spectra is consistent to the effective magnetic moment,
$\mu_{eff}=5.8$ $\mu_{B}$/f.u., obtained from the $H_{dc}=500$ Oe $M(T)$ curve at the paramagnetic state (see below). Applying the usual equation for
systems with two or more different magnetic ions \cite{Niebieskikwiat}
\begin{equation}
\mu= \sqrt{{\mu_1}^2 + {\mu_2}^2 + {\mu_3}^2 +...} \label{EqMu}
\end{equation}
and using the standard magnetic moments of Ir$^{4+}$ (1.73 $\mu_{B}$), HS Co$^{2+}$ (5.2 $\mu_{B}$) and HS Co$^{3+}$ (5.48 $\mu_{B}$), yields
\begin{equation}
\mu = \sqrt{0.7(5.2)^2 + 0.3(5.48)^2 + (1.73)^2} = 5.6 \mu_{B}/f.u..
\end{equation}

The difference to the experimental value may be related to spin-orbit coupling (SOC) on Co ions, as already reported for similar compounds
\cite{Kolchinskaya,Raveau}.

\subsection{ac and dc magnetization vs. $T$}

\begin{figure}
\begin{center}
\includegraphics[width=0.4 \textwidth]{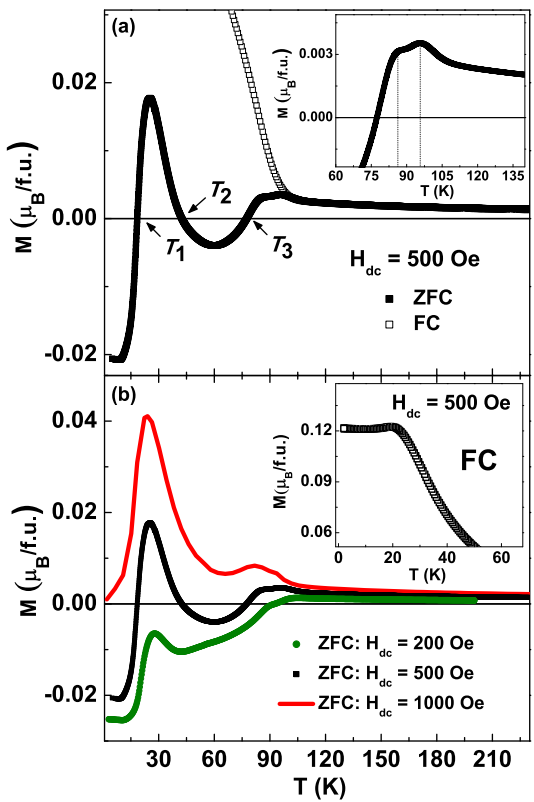}
\end{center}
\caption{(a) ZFC and FC magnetization as a function of $T$ for La$_{1.5}$Ca$_{0.5}$CoIrO$_{6}$ at $H_{dc}=500$ Oe.  The inset shows the magnified view of
the AFM and FM transitions of Co ion. (b) ZFC magnetization as a function of $T$ at $H_{dc}=200$ Oe (solid circle), $H_{dc}=500$ Oe (solid square), and
$H_{dc}=1000$ Oe (solid line). The inset shows the $H_{dc}=500$ Oe FC curve at low-$T$.}
\label{FigMxT}
\end{figure}

Fig. \ref{FigMxT}(a) presents the ZFC-FC magnetization curves for $H_{dc}=500$ Oe, were two peaks can be clearly observed. The inset shows a magnified view
of a stretch of the curve, where are evidenced two anomalies at $T\simeq97$ and $T\simeq86$ K, associated to the magnetic ordering of the AFM and FM phases
of Co ions, respectively, as it will be addressed later. The lower-$T$ cusp at $\simeq27$ K is due to a SG-like behavior, indicating a RSG phenomena. Since
each peak is associated to a distinct mechanism, if an appropriate $H_{dc}$ is applied in the opposite direction of the material's spontaneous
magnetization ($M_{sp}$), the curve can be shifted down and display up to three reversals of its magnetic moment.

This scenario can be understood in terms of the AFM coupling of the Co ions that are located at their predicted site, resulting in the frustration of
Ir$^{4+}$ magnetic moment, and the development of a FM component caused by the ASD. Studies of XANES on Co $K$- and Ir $L_{2}$-,$L_{3}$-edges indicate a Co
mixed valence on La$_{1.5}$Ca$_{0.5}$CoIrO$_{6}$ (See Supplemental Material \cite{Supp}). Also, the XPS measurement discussed above suggests a proportion
$\sim$70\%/30\% of high spin Co$^{2+}$/Co$^{3+}$. Ir maintains 4+ valence by La$^{3+}$ to Ca$^{2+}$ substitution. Reports of neutron powder diffraction
studies on La$_{2-x}$Sr$_{x}$CoIrO$_{6}$ \cite{Kolchinskaya} and several closely related compounds (\textit{e.g.} La$_{2-x}$Sr$_{x}$CoRuO$_{6}$
\cite{Dlouha} and LaBaCoIrO$_{6}$ \cite{Battle}) revealed AFM coupling of Co ions on these double perovskites, as predicted by the
Goodenough-Kanamori-Anderson (GKA) rules \cite{Goodenough}. Moreover, our electronic structure calculation results (see Supplemental Material \cite{Supp})
indicate the AFM ordering as the most stable structure for La$_{1.5}$Ca$_{0.5}$CoIrO$_{6}$. Therefore, this spin orientation can be assumed for the
majority of Co ions. Consequently, the Ir$^{4+}$ located in between the AFM coupled Co$^{2+}$ ions might be frustrated. However, the system exhibit 9\% of
ASD, which together with the 30\% of Co$^{3+}$ leads to other nearest neighbor interactions such as Co$^{2+}$--O--Co$^{3+}$, Co$^{3+}$--O--Co$^{3+}$ and
Ir$^{4+}$--O--Ir$^{4+}$. The latter two couplings are AFM, as predicted by the GKA rules \cite{Seo,Kim,Goodenough}, but Co$^{2+}$--O--Co$^{3+}$ coupling is
expected to be a short range FM interaction via double-exchange mechanism \cite{Raveau,Seikh,Singh}.

The resulting magnetic moment of La$_{1.5}$Ca$_{0.5}$CoIrO$_{6}$ can be estimated by the following argument. The main phase consists of AFM coupled
Co$^{2+}$ and frustrated Ir ions. Hence there is no contribution from the main phase to the resulting magnetization, as well as the AFM
Co$^{3+}$--O--Co$^{3+}$ and Ir$^{4+}$--O--Ir$^{4+}$ couplings. To calculate the magnetization per formula unit, it must be taken into account that from the
30\% of the Co$^{3+}$ present in the system, in 9\% it will permute to Ir site, giving rise to the FM Co$^{2+}$--O--Co$^{3+}$ interaction, which is the
only from the interactions discussed above that is expected to contribute to the net magnetization. Using the standard magnetic moments of HS Co$^{2+}$ and
Co$^{3+}$, the FM contribution to the resulting magnetization can be estimated to be
\begin{equation}
\begin{split}
M& = 0.3 \times 0.09(0.5M_{Co^{3+}}+0.5M_{Co^{2+}}) \\
 & = 0.027(0.5\times5.48+0.5\times5.2) \\
 & = 0.14 \mu_{B}/f.u..
\end{split}
\end{equation}

This result is very close to the low-$T$ value of the FC magnetization observed on inset of Fig. \ref{FigMxT}(b), 0.12 $\mu_{B}$/f.u., and to the remanent
magnetization of $M(H)$ curve at 2 K ($M_{R}=0.13$ $\mu_{B}$/f.u., see Fig. \ref{FigZEB}). It is also close to the total magnetic moment obtained from band
structure calculation, $M=0.15$ $\mu_{B}$/f.u. (see Supplemental Material \cite{Supp}). Surely, the above calculation is only an estimate, since it is not
possible to establish the exact individual contribution of Co$^{2+}$ and Co$^{3+}$ magnetic moments. Also, the Co$^{2+}$/Co$^{3+}$ proportion may vary a bit
from that obtained from XPS, leading to the discrepancy between the observed and calculated values. The slightly larger experimental values may be also
related to the contribution of the frozen SG ions, which were not taken into account in the calculation. The above approximation only takes into account
the linear exchange of Ising moments. In a 3D system, other exchange pathways may play a role, resulting in a more complex picture, but also with competing
FM and AFM interactions and perhaps even canted spins \cite{Goodenough,Greedan}. Hence, also for a 3D magnetic model, one would expect Ir$^{4+}$
frustration. Frustration, along with the magnetic segregation due to ASD, fit very well the material's magnetic behavior.

The magnetic interactions discussed above can explain the curve of Fig. \ref{FigMxT}(a) as the following. After ZFC, at low-$T$, there is only the $M_{sp}$
related to the Co interactions, since the Ir moments are frustrated. The system's $M_{sp}$ is here always chosen as opposite (negative) to $H_{dc}$
direction (positive). Thus, applying $H_{dc}$ in the opposite direction of $M_{sp}$, and ascending to higher-$T$ there is the first compensation
temperature at $T_{1}\simeq18$ K, due to the SG-spins alignment to $H_{dc}$ direction. By increasing $T$ there is the decrease of the SG correlation
length, resulting in the second magnetization reversal at $T_{2}\simeq42$ K. With the further enhancement of the thermal energy the Co-FM phase can achieve
positive magnetization, and there is the third compensation temperature at $T_{3}\simeq78$ K. Finally, there are the transitions of the Co$^{2+}$-AFM and
Co$^{2+}$/Co$^{3+}$-FM phases to the paramagnetic state.

This behavior is closely related to the dynamics of the spin clusters, which are strongly dependent on $H_{dc}$ and measurement time. For different times
of measurement the compensation temperatures can vary. For a larger $H_{dc}$, the system achieves positive magnetization values already at low-$T$, and
there is no magnetic reversal, as can be observed on Fig. \ref{FigMxT}(b) for the curve measured under $H_{dc}=1000$ Oe. On the other hand, for small
$H_{dc}$ the SG peak is not large enough to achieve the positive magnetization and the system undergoes only to one compensation temperature, $T_{3}$. For
$H_{dc}=200$ Oe, one have $T_{3}=92$ K, as shown in Fig. \ref{FigMxT}(b). Another important result obtained from Fig. \ref{FigMxT}(b) is that the low $T$
peak maxima shifts to lower-$T$ as $H_{dc}$ increases. For $H_{dc}=200$, 500 and 1000 Oe the peak maxima are located at 27.8, 25.4 and 23.9 K respectively.
This is an expected feature of a SG-like material, as will be discussed next.

\begin{figure}
\begin{center}
\includegraphics[width=0.5 \textwidth]{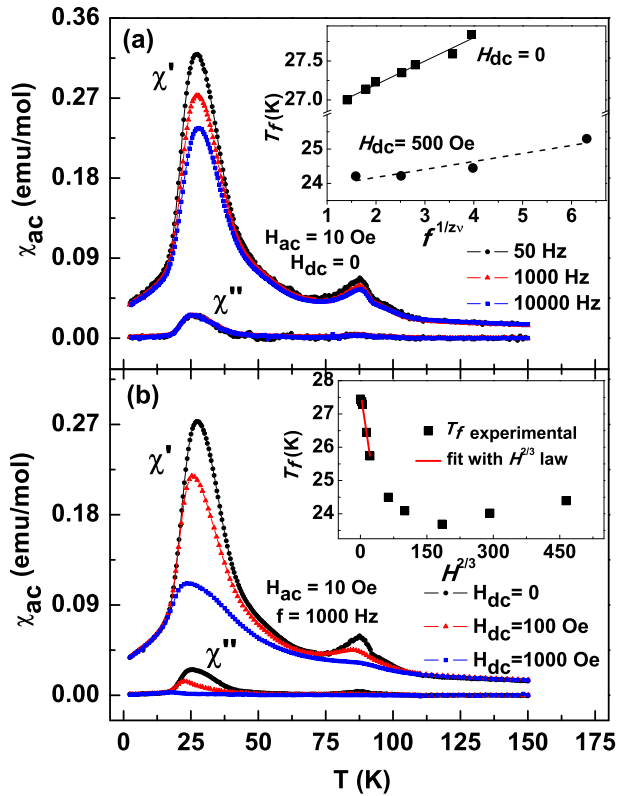}
\end{center}
\caption{(a) $\chi$'$_{ac}$ and $\chi$''$_{ac}$ as a function of $T$ at various frequencies. The inset shows $T_{f}$ as a function of frequency for
$H_{dc}$=0 and $H_{dc}$=500 Oe, obtained from $\chi$'$_{ac}$. The lines are best fits to the power law $T_{f}$=$T_{sg}$[1+($\tau$$_{0}$$f$)$^{1/z\nu}$)].
(b) $\chi$'$_{ac}$ and $\chi$''$_{ac}$ vs. $T$ at $f$=1000 Hz for various $H_{dc}$. Inset shows $T_{f}$ for different applied $H_{dc}$.}
\label{FigChiAC}
\end{figure}

The ac magnetic susceptibility ($\chi_{ac}$) was measured as a function of $T$ for seven frequencies in the range 10-10$^{4}$ Hz. Fig. \ref{FigChiAC}(a)
shows the real ($\chi$'$_{ac}$) and imaginary ($\chi$''$_{ac}$) parts of susceptibility for some selected frequencies. For $\chi$'$_{ac}$ it was observed
for the low-$T$ peak ($T_{f}$) a systematic  shift of $T_{f}$ to higher-$T$ with the increase of driving frequency, where $T_{f}$ shift from 27 K at 10 Hz
to 27.84 K at 10$^{4}$ Hz. It was also observed a decrease of the peak amplitude with increasing frequency. Both results are characteristic of SG-like
materials. On the other hand, for the $T\simeq90$ K only a small amplitude decrease was observed at higher frequencies, but no measurable shift, which
characterizes ordinary magnetic transitions.

The frequency-dependent data turn out to be well described by the conventional critical slowing down model of the dynamic scaling theory
\cite{Mydosh,Souletie,Hohenberg}, which predicts a power law
\begin{equation}
\frac{\tau}{\tau_{0}}=\left[\dfrac{(T_{f} - T_{sg})}{T_{sg}}\right]^{-z\nu} \label{Eq1}
\end{equation}
where $\tau$ is the relaxation time corresponding to the measured frequency, $\tau$$_{0}$ is the characteristic relaxation time of spin flip, $T_{sg}$ is
the SG transition temperature (as frequency tends to zero), $z$ is the dynamical critical exponent and $\nu$ is the critical exponent of the correlation
length. The solid line in the inset of Fig. \ref{FigChiAC}(a) represents the best fit to the power law divergence, that yields $T_{sg}=26.6$ K,
$\tau_{0}=1.5\times 10^{-13}$ s and $z\nu=6.5$. These results are in the realm of conventional SG phases.

A criterion that is often used to compare the frequency dependence of $T_{f}$ in different SG systems is to compare the relative shift in $T_{f}$ per
decade of frequency
\begin{equation}
\delta T_{f}=\frac{\bigtriangleup T_{f}}{T_{f} \bigtriangleup(log f)}. \label{Eq2}
\end{equation}

For  La$_{1.5}$Ca$_{0.5}$CoIrO$_{6}$ we found $\delta T_{f}\simeq0.008$, which is within the range usually found for conventional SG ($\delta
T_{f}\lesssim0.01$). For superparamagnets the usual value is $\delta T_{f}\gtrsim0.1$, while for cluster glasses (CG) it has intermediate values between
canonical SG and superparamagnets \cite{Mydosh,Souletie,Anand2,Malinowski}.

The SG cusps on Fig. \ref{FigChiAC}(a) are broader than usually observed for canonical SG. Since La$_{1.5}$Ca$_{0.5}$CoIrO$_{6}$ is a RSG material, this
may be due to the internal molecular field resulting from Co moments. In order to verify the field effect on the magnetization it was measured $\chi_{ac}$
with $H_{dc}=500$ Oe. Inset of Fig. \ref{FigChiAC}(a) shows that, for this $H_{dc}$, $T_{f}$ also shifts to higher-$T$ with increasing frequency. The
dashed line is the fit to the power law, yielding $T_{sg}^{500Oe}=23.7$ K, $z\nu=5$, $f_{0}=10^{10}$ Hz. With $H_{dc}=500$ Oe, the relative shift in
$T_{f}$ (Eq. \ref{Eq2}) increases to $\delta T_{f}\simeq0.01$. All these results are compatible to those usually reported for CG materials. Hence, $H_{dc}$
induces the increase of the correlation length of the spin clusters, \textit{i.e.}, there is a transition from SG to CG in the system.

Fig. \ref{FigChiAC}(b) shows $\chi$'$_{ac}$ and $\chi$''$_{ac}$ measurements for fixed $f=1000$ Hz and $H_{ac}=10$ Oe, but different $H_{dc}$. As expected,
the SG peak is smeared out and shifts to lower-$T$ with increasing $H_{dc}$ \cite{Mydosh}. The inset shows that $T_{f}(H)$ reasonably follows the $H^{2/3}$
Almeida-Thouless relation \cite{Almeida} for $H_{dc}\leq100$ Oe, and the curve's slope changes for higher fields. Since La$_{1.5}$Ca$_{0.5}$CoIrO$_{6}$ is
a RSG compound, $H_{dc}$ have its effect on the underlying Co ions, inducing the transition from conventional SG to CG. This result may bring important
insights about the effect of strong $H_{dc}$ on the frozen spins and also on the limit of validity of the $H^{2/3}$ relation for a RSG. $H_{dc}$ seems to
remove the criticality of the transition, yet it does not fully prevent the formation of the frozen state \cite{Mydosh}.

In contrast to $\chi$'$_{ac}$, for $\chi$''$_{ac}$ there was a small non-monotonic variation of the peak position and amplitude. This is an unconventional
behavior for a SG-like material, and it was not found a reasonable simple explanation for it. It was already reported for the ternary intermetallic
CeRhSn$_{3}$ a shift toward lower-$T$ with increasing frequency \cite{Anand2,Anand1}. But differently than for La$_{1.5}$Ca$_{0.5}$CoIrO$_{6}$, the shift
was on $\chi$', while $\chi$'' goes to higher-$T$ with increasing frequency. Here is important to stress that, despite the fact our measurements were
carefully taken, the overall smaller and more noisy $\chi$''$_{ac}$ makes it more difficult to precisely determine the $T_{f}$ position, specially for
lower frequencies.

\begin{figure}
\begin{center}
\includegraphics[width=0.5 \textwidth]{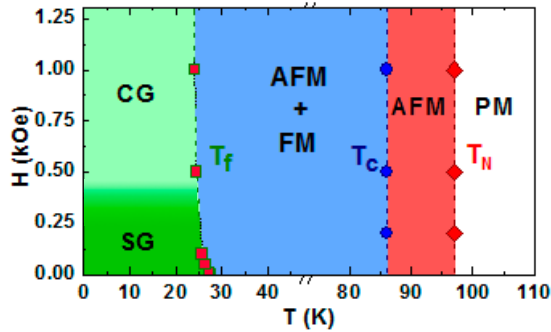}
\end{center}
\caption{Temperature-field phase diagram showing the N\'{e}el ($T_N$), critical ($T_C$) and freezing ($T_f$) temperatures of
La$_{1.5}$Ca$_{0.5}$CoIrO$_{6}$. The lines are guides for the eye.}
\label{FigPD}
\end{figure}

The complex magnetic behavior of La$_{1.5}$Ca$_{0.5}$CoIrO$_{6}$ can be summarized in a rich phase diagram, showing its conventional and SG-like states.
Fig. \ref{FigPD} displays the field-temperature phase diagram for the compound, where can be observed its N\'{e}el temperature ($T_N$) at $T_{N}\simeq97$
K, its critical temperature ($T_C$) at $T_{C}\simeq86$ K, and its freezing temperature ($T_f$) at $T_{f}\simeq26.6$ K. An important result observed is
that, depending on $H_{dc}$, at low-$T$ the material can behave as a conventional spin glass or as a cluster glass.

\subsection{Exchange bias}

\begin{figure}
\begin{center}
\includegraphics[width=0.35 \textwidth]{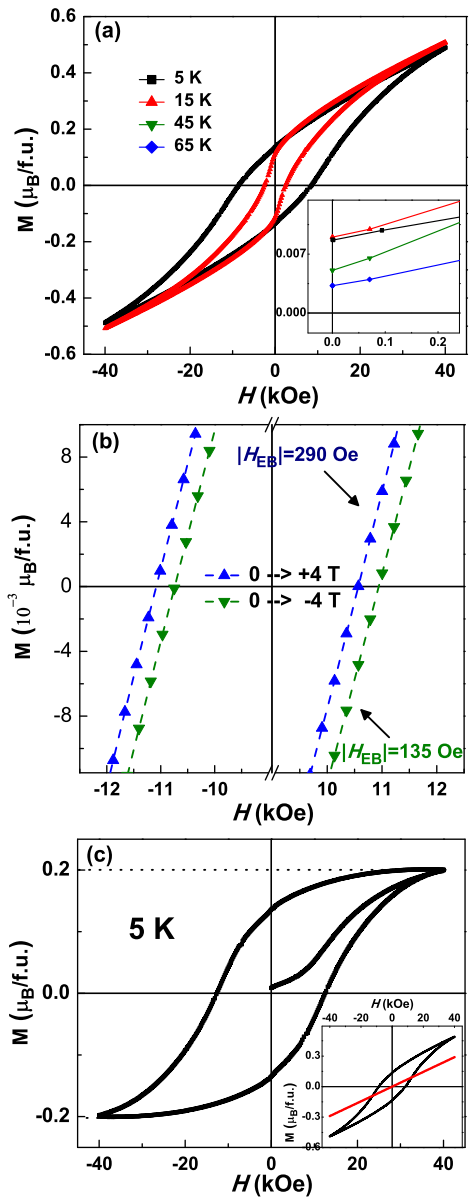}
\end{center}
\caption{(a) ZFC $M(H)$ loops at 5 and 15 K. The inset shows a magnified view of the initial magnetization values at zero field. (b) ZFC $M(H)$ loops at 2
K performed as $0\rightarrow4$ T$\rightarrow-4$T$\rightarrow4$ T and $0\rightarrow-4$ T$\rightarrow4$ T$\rightarrow-4$ T. (c) FM + SG contributions to the
$M(H)$ loop at $T=5$ K. The doted lines are guides for the eye. The inset shows the original curve and the linear AFM contribution (see text).}
\label{FigMxH}
\end{figure}

$M(H)$ curves for ordinary FM and FIM materials usually exhibit hysteretic behavior with coercive field due to the blocking of the domain wall motion. In
SG-like materials irreversibility can also be observed arising out of anisotropy \cite{Mydosh,Mukherjee,De}. Hence, in a RSG system a large anisotropic
coercivity can be expected due to the combined action of FM, AFM and SG phases. The ZFC $M(H)$ measurements were performed for several $T$ using a
systematic protocol detailed on Supplemental Material \cite{Supp}, and two representative curves are displayed in Fig. \ref{FigMxH}(a), where one can see a
small increase in the magnetization from 5 to 15 K. The increase of thermal energy results in an enhancement of the SG alignment to the field direction,
yielding in a larger magnetization. However, due to thermal energy, these spins can flip to the field direction, leading to the observed decrease of
coercivity. It is also important to note that the system exhibit a non-negligible $M_{sp}$ at zero field [inset of Fig. \ref{FigMxH}(a)]. This $M_{sp}$
shows a systematic evolution with $T$, and plays an important role on the process of pinning the spins, as it will be discussed next.

Here we define the EB field as $H_{EB}=|H_{+}+H_{-}|/2$, where $H_{+}$ and $H_{-}$ represent the right and left field values of the $M(H)$ loop at the
$M=0$ axis, respectively. The effective coercive field is $H_{C}=|H_{+}-H_{-}|/2$. Usually, the EB effect is achieved when the system is cooled in the
presence of non-zero $H_{dc}$. Interestingly, for La$_{1.5}$Ca$_{0.5}$CoIrO$_{6}$ a non-negligible shift of the hysteresis loop is observed even when the
system is cooled in zero field. This spontaneous EB effect, also called zero field cooled EB (ZEB), was recently reported for distinct systems such as
Mn$_{2}$PtGa \cite{Nayak} and Ni-Mn-In \cite{Wang} alloys, and the nanocomposite BiFeO$_{3}$-Bi$_{2}$Fe$_{4}$O$_{4}$ \cite{Maity}. But here we have found,
to the best of our knowledge, the first example of a material to have this phenomenon clearly related to three distinct magnetic phases, namely FM, AFM and
SG. At 2 K La$_{1.5}$Ca$_{0.5}$CoIrO$_{6}$ exhibit a negative shift $H_{ZEB}\simeq290$ Oe. In order to verify this effect we have measured $M(H)$ with the
initial $H_{dc}$ in the opposite direction. As can be observed on Fig. \ref{FigMxH}(b), the curve exhibits a positive shift $H_{ZEB}\simeq135$ Oe. The
shift in the opposite direction is an expected behavior of a EB system, hence a clear evidence that this result is intrinsic of the material. The fact that
$H_{ZEB}$ is different depending on the direction of the initial magnetization process is an indicative that the internal $M_{sp}$ plays an important role
in the pinning of the spins.

La$_{1.5}$Ca$_{0.5}$CoIrO$_{6}$ presents FM and SG phases incorporated to an AFM matrix. Due to its predominant AFM phase, it does not saturate even at a
large field of 9 T. As will be discussed next, the ZEB effect here observed results from the a delicate exchange interaction on the interfaces of the
AFM/SG phases to the minor FM phase. Hence, for large enough applied fields the pinned spins at the interface may flip to the field direction, reducing the
effect. This is actually what is observed for fields larger than 4 T. For instance, for a maximum applied field ($H_{m}$) of 9 T the EB effect is reduced
to $H_{ZEB}$=40 Oe. In order to evidence the FM contribution to the $M(H)$ curves, the AFM contribution was subtracted from the loops. The linear curve
representing the AFM phase was obtained from the fit of the loops at high fields, which was extrapolated to the whole field range and then subtracted from
the loop. The resulting curve obtained for $T=5$ K is displayed in Fig. \ref{FigMxH}(c). It is almost symmetric in respect to the $M$ axis and displays the
same $H_{ZEB}$ as that obtained from the original curve. It must be mentioned that the resulting curves contemplate both the contributions of the FM and SG
phases, \textit{i.e.}, it is not possible to separate these phases on the curves.

Despite the ZEB being an effect only recently reported, the conventional exchange-bias (CEB) is a well known phenomenon encountered in systems containing
interfaces between distinct magnetic phases, being most likely found in FM-AFM systems. But a shift of the magnetization hysteresis loops along the field
axis can be also observed in situations not related to EB effect. In a conventional FM material, if a minor $M(H)$ loop is measured, \textit{i.e.}, an
$M(H)$ with the maximum applied field not large enough to the system achieve the magnetic saturation, it can exhibit a shift along the field axis similar
to that observed in EB systems. This is in general related to the incomplete magnetic reversion of the system \cite{Becker,Benda}. However, differently
than observed in EB materials, these minor loops also exhibit a large shift along the magnetization axis, and in general the loops are not closed at large
fields \cite{Pi,Klein}. As Fig. \ref{FigMxH} shows, $M(H)$ of La$_{1.5}$Ca$_{0.5}$CoIrO$_{6}$ is a closed loop with a very small shift along the vertical
axis due to the pinned spins, as expected. It is a indicative that ZEB reported here is not due to a minor loop hysteresis. In order to reinforce the
difference between the curves here observed from those obtained from minor loops, we compare in the Supplemental Material the results here described with
that obtained from true minor loops \cite{Supp}.

\begin{figure}
\begin{center}
\includegraphics[width=0.4 \textwidth]{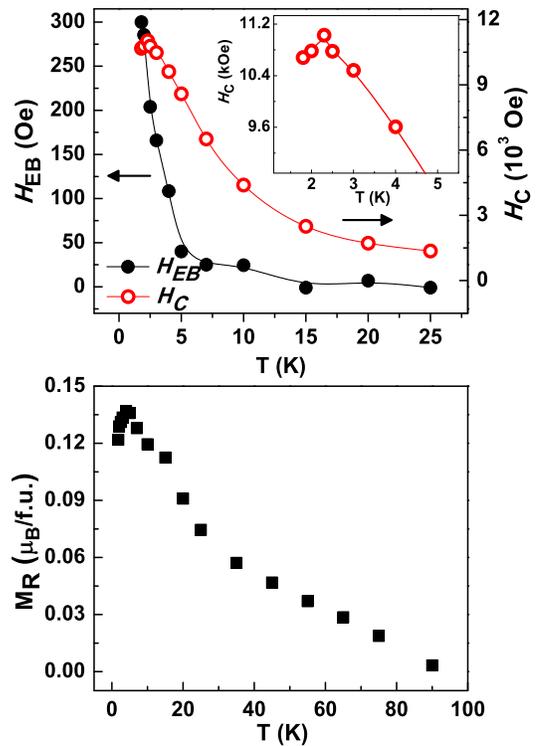}
\end{center}
\caption{(a) $H_{ZEB}$ and $H_{C}$ evolution with $T$. The inset shows a magnified view of the low-$T$ anomaly on $H_{C}$.The lines are guides for the eye.
(b) $M_{R}$ dependency with $T$.}
\label{FigZEB}
\end{figure}

The evolution of $H_{ZEB}$ and $H_{C}$ with $T$ are displayed on Fig. \ref{FigZEB}(a). One can observe that the ZEB effect can only be achieved below
$T_{f}$. Moreover, it rapidly decreases with the enhancement of the thermal energy. Usually, a decrease in the magnetic unidirectional anisotropy (UA) is
associated with an increase of the coercivity \cite{Nogues}. Hence, the low-$T$ anomaly observed for $H_{C}$ on inset of Fig. \ref{FigZEB}(a) is another
expected feature of EB systems.

Different mechanisms are invoked to explain the ZEB for distinct materials. For instance, for Ni-Mn-In it is proposed that the UA is formed at the
interface between different magnetic phases during the initial magnetization process of $M(H)$ curves \cite{Wang}. On the other hand, for
BiFeO$_{3}$-Bi$_{2}$Fe$_{4}$O$_{4}$ it is proposed that glassy moment at the interface between FM-AFM phases causes the EB effect \cite{Maity}. Despite the
distinct mechanisms claimed to be the responsible for the ZEB effect on each compound, they all have in common the RSG behavior. Here we conjecture that
the internal molecular field plays an important role on the ZEB effect. We propose the following mechanism to explain the ZEB effect on
La$_{1.5}$Ca$_{0.5}$CoIrO$_{6}$. The internal field due to the FM phase have its impact on AFM and SG phases. It affects the correlation length of the SG
clusters, favoring the freezing of the glassy spins in the same direction, resulting in a spontaneous UA. The ZEB is enhanced by $H_{dc}$ during the
initial magnetization process of the $M(H)$ loop. The field induces the increase of the internal interaction of FM domains, leading to the growth of the
spin clusters. After the removal of $H_{dc}$, the spins at the grain interfaces are pinned, resulting in a stable magnetic phase with UA at low-$T$. As $T$
increases the pinned spins can easier flip to the field direction due to the enhanced thermal energy. This leads to the reduction of $H_{ZEB}$ and the
correlated increase of $H_{C}$ observed on Fig. \ref{FigZEB}(a).

The same scenario can explain the system's remanent magnetization, $M_{R}=|M^{+}_{R}-M^{-}_{R}|/2$, where $M^{+}_{R}$ and $M^{-}_{R}$ are the positive and
negative values of the magnetization at zero field. Fig. \ref{FigZEB}(b) displays the $M_{R}$ evolution with $T$. On going from low to high-$T$, first
there is an increase of $M_{R}$ due to the thermally activated movement of the spins to the field direction. Then going to higher-$T$ there is the
continuous decrease of $M_{R}$ until it vanishes at the paramagnetic sate. It is important to observe that despite the fact the FM phase orders at $\sim90$
K, the ZEB effect only occurs below $T_{f}$. This, together with the fact that at low-$T$ $M_{R}$ and $H_{C}$ initially increase while $H_{EB}$ decreases
on increasing $T$, are other evidences that the ZEB observed is strongly related to the RSG state and is not due to some minor hysteresis loop of the FM
phase.

\begin{figure}
\begin{center}
\includegraphics[width=0.4 \textwidth]{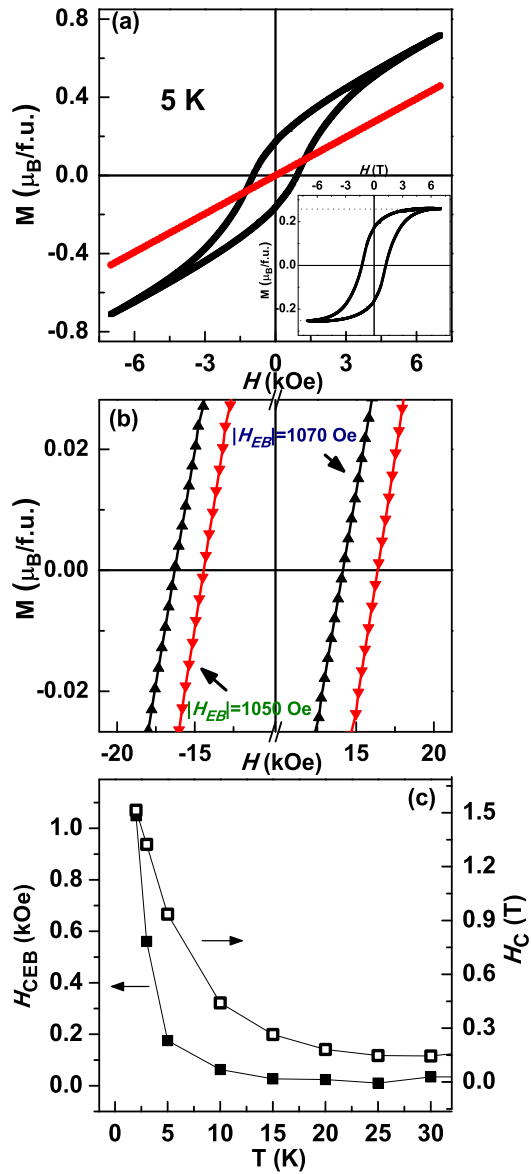}
\end{center}
\caption{(a) $M(H)$ loop at 5 K after cooling the system with $H_{FC}=3$ T. The inset shows the deconvolute curve (see text). (b) Magnified view of the
curves measured at 2 K after the sample being field cooled with $H_{FC}\pm3$ T. (c) $H_{CEB}$ and $H_{C}$ evolution with $T$.}
\label{FigCEB}
\end{figure}

As addressed above, a significant evidence that the observed ZEB effect on La$_{1.5}$Ca$_{0.5}$CoIrO$_{6}$ is not due to some experimental artifact is the
fact that the $M(H)$ loops shift to opposite directions depending on the initial applied field be positive or negative. If there were remanent current on
the magnet due to an trapped flux or any other reason, both shifts should be expected to be in the same direction. This inversion of $H_{EB}$ depending on
the initial field value is also observed when the CEB effect is measured, \textit{i.e.}, when the isothermal $M(H)$ curve is measured after the system
being field cooled. Fig. \ref{FigCEB}(a) displays the $M(H)$ loop with $H_{m}=7$ T, after the sample being cooled in the presence of $H_{FC}=3$ T. The AFM
contribution is also displayed, and the inset shows the deconvoluted curve, \textit{i.e.}, the resulting curve when the AFM one is subtracted. On Fig.
\ref{FigCEB}(b) is shown a magnified view of the 2 K loops after the sample being cooled with $H_{FC}=\pm3$ T. Differently than for the ZEB curves, for CEB
the two curves are nearly symmetrically displaced with respect to the magnetization axis. Here the $\pm3$ T cooling field is strong enough to flip the
spontaneous magnetization, and the effect of the internal field becomes negligible.

Fig. \ref{FigCEB}(c) shows the temperature dependence of $H_{CEB}$ and $H_{C}$. The CEB evolution is similar to ZEB. Above $T_{f}$ the $H_{CEB}$ becomes
negligible. It shows the importance of the SG phase to the EB observed on La$_{1.5}$Ca$_{0.5}$CoIrO$_{6}$. It can be also observed that the EB effect is
greatly enhanced when the system is field cooled. At 2 K one have $H_{CEB}$$\simeq1070$ Oe. On the FC procedure, the pinning of the SG spins is favored
already from above $T_{f}$ down to low-$T$, and these spins get freezed on the field direction.

\begin{figure}
\begin{center}
\includegraphics[width=0.45 \textwidth]{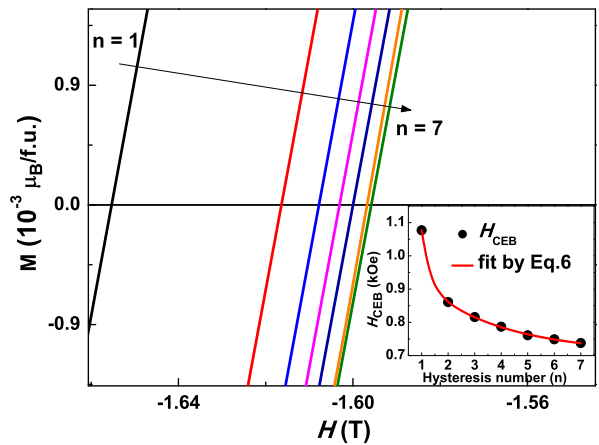}
\end{center}
\caption{Training effect of CEB at 2 K. The inset shows $H_{CEB}$ as a function of the hystereis number (n). The solid line represents the fitting of the
experimental data to Eq. \ref{EqTE}.}
\label{FigTE}
\end{figure}

In EB systems, repeating the $M(H)$ loop may lead to relaxation of uncompensated spin configuration at the interface. Consequently, $H_{EB}$ depends on the
number of consecutive hysteresis loops measured. This property is called training effect. For La$_{1.5}$Ca$_{0.5}$CoIrO$_{6}$ this behavior was
investigated in both ZEB and CEB cases. For CEB, 7 consecutive loops were measured at 2 K, after cooling the sample in the presence of $H_{FC}=3$ T. Fig.
\ref{FigTE} shows a detailed view of the 7 loops close to the $M=0$ axis. The arrow indicates a systematic evolution of the curves. The dependence of
$H_{CEB}$ on the number of repeating cycles ($n$) is shown on inset. As can be observed, $H_{CEB}$ decreases monotonically with the increase in $n$,
indicating spin rearrangement at the interface. The $n$ dependence of $H_{CEB}$ can be fit to a model considering the contribution of both the frozen spins
and the uncompensated rotatable spins at the interface \cite{Maity,Mishra}
\begin{equation}
{H}^n_{CEB} = H^{\infty}_{CEB} + A_{f}e^{(-n/P_{f})} + A_{r}e^{(-n/P_{r})}, \label{EqTE}
\end{equation}
where $f$ and $r$ denote the frozen and rotatable spin components respectively. Eq. \ref{EqTE} fits the data very well, for $H^{\infty}_{CEB}=708$ Oe,
$A_{f}=4627$ Oe, $P_{f}=0.3$, $A_{r}=282$ Oe, $P_{3}=3.1$. The fact $A_{f}\textgreater A_{r}$ indicates the importance of the SG phase to the EB effect,
and $P_{r}\textgreater P_{f}$ suggests that the rotatable spins rearrange faster than the frozen ones. For the ZEB mode, it was observed only a very small
decrease of $H_{ZEB}$ from first to second loop, thereafter the system exhibits only negligible variation. This indicates that after the first cycle, the
frozen and uncompensated spins became quite stable at the interfaces.

\section{Conclusions and Outlook}

In conclusion, we have shown that  La$_{1.5}$Ca$_{0.5}$CoIrO$_{6}$ is a RSG-like material, in which there are two magnetic orderings of the AFM and FM
phases of Co ions at $T_{N}=97$ K and $T_{C}=86$ K, respectively, and a SG-like transition at $T_{sg}=26.6$ K. The frequency dependence of $T_{f}$ obtained
from $\chi$'$_{ac}$ follows the power law of the dynamical scaling theory. Regarding the $\chi$$_{ac}$ measurements with applied $H_{dc}$ fields, the
system transits from conventional spin glass to cluster glass with increasing $H_{dc}$. The coexistence of conventional and glassy magnetic states leads to
an exotic magnetic behavior in the ZFC $T$-dependence of magnetization curve, in which the system can undergo three magnetic reversals. Magnetization as a
function of $H_{dc}$ suggest a ZEB effect at low-$T$, related to the FM-AFM-SG interfaces. When the sample is cooled in the presence of an applied magnetic
field, the EB effect is enhanced. XPS, XANES, XMCD and electronic structure calculations results corroborate our argument that the magnetization reversals
and the EB effect can be both understood in terms of the same underlying mechanism, \textit{i.e.}, are consequences of the Ir magnetic frustration caused
by the competing interactions with its neighboring Co ions. To verify these and other conjectures discussed in the text, other techniques such as neutron
scattering, torque magnetometry and electronic spin resonance are necessary.

\begin{acknowledgements}
This work was supported by CNPq, FAPERJ, FAPESP and CAPES (Brazil). We thank E. Granado for the helpful discussions. F. Stavale thanks the Surface and
Nanostructures Multiuser Lab at CBPF and the MPG partnergroup programm. LNLS is acknowledged for concession of beam time.
\end{acknowledgements}


\newpage



\renewcommand{\figurename}{Supplemental FIG.}
\renewcommand{\tablename}{Supplemental TABLE}
\setcounter{figure}{0}

\section*{Supplementary Material: ``Compensation temperatures and exchange bias in La$_{1.5}$Ca$_{0.5}$CoIrO$_{6}$"}

\subsection*{X-ray near edge structure (XANES)}

Room temperature x-ray near edge structure (XANES) measurements were performed in the dispersive x-ray absorption (DXAS) beam line at the Brazilian Synchrotron Light Laboratory (LNLS) \cite{s_dxas}. The edge step normalization of the data was performed after a linear pre-edge subtraction and the regression of a quadratic polynomial beyond the edge, using the software ATHENA \cite{s_athena}.

\begin{figure}
\begin{center}
\includegraphics[width=0.4 \textwidth]{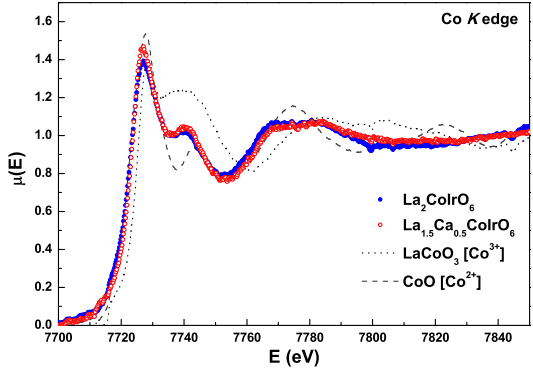}
\vspace{-0.8 cm}
\end{center}
\caption{Normalized Co $K$ edge XANES spectra of La$_{2}$CoIrO$_{6}$ and La$_{1.5}$Ca$_{0.5}$CoIrO$_{6}$ at room temperature. The references LaCoO$_3$ and CoO XANES spectrum is also shown for comparison.}
\label{FigS3}
\end{figure}

The normalized Co $K$-edge XANES spectrum $\mu(E)$ of La$_{2}$CoIrO$_{6}$ and La$_{1.5}$Ca$_{0.5}$CoIrO$_{6}$ are given in Fig. \ref{FigS3}. The references LaCoO$_3$ and CoO XANES spectrum are also shown for comparison. Interestingly, our results indicate the presence of Co$^{3+}$ already at the parent compound, which was also observed on the XPS. The small shift to higher energies obseved for La$_{1.5}$Ca$_{0.5}$CoIrO$_{6}$ suggests an increase in the proportion of Co$^{3+}$ due to Ca$^{2+}$ substitution on La$^{3+}$ site.

\begin{figure}
\begin{center}
\includegraphics[width=0.4 \textwidth]{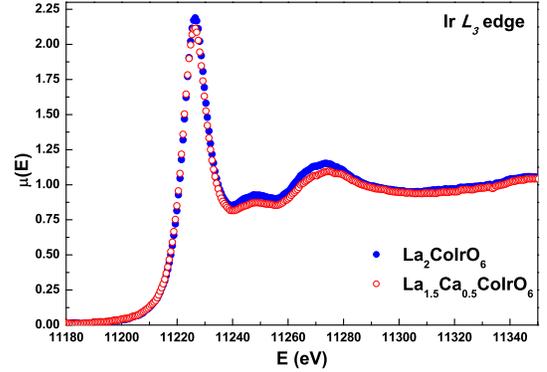}
\vspace{-0.8 cm}
\end{center}
\caption{Normalized Ir $L_3$ edge XANES spectra of La$_{2}$CoIrO$_{6}$ and La$_{1.5}$Ca$_{0.5}$CoIrO$_{6}$ at room temperature.}
\label{FigS4}
\end{figure}

The normalized Ir $L_3$-edge XANES spectrum $\mu(E)$ of La$_{2}$CoIrO$_{6}$ and La$_{1.5}$Ca$_{0.5}$CoIrO$_{6}$ are given in Fig. \ref{FigS4}. Our XANES measurements indicate no appreciable change in the Ir valence by calcium doping. This is in agreement with the magnetometry results and to the XPS spectra at Ir 4$f$ region (not shown). However, it can be observed a small variation on the wavelength intensity from La$_{2}$CoIrO$_{6}$ to La$_{1.5}$Ca$_{0.5}$CoIrO$_{6}$. Although our results indicate a majority of Ir$^{4+}$, the possible valence mixing in Ir sublattice can not be discarded. Especially if we assume the oxygen content to be nearly stoichiometric, it should be expected a few percentage of Ir$^{5+}$ ions. Recent reports indicate long-range magnetic order in Ir$^{5+}$-based double-perovskites \cite{s_Dey}, hence it could contribute to the magnetic behavior observed for La$_{1.5}$Ca$_{0.5}$CoIrO$_{6}$.

\subsection*{X-ray magnetic circular dichroism (XMCD)}

X-ray magnetic circular dichroism (XMCD) measurements at the Ir $L_{2,3}$ edges were performed in the DXAS beam line at LNLS \cite{s_dxas}, with a calculated degree of circular polarization of $\sim75$\%. A rotative permanent magnet applied 0.9 T at the sample, both parallel and antiparallel to the x-ray beam direction.

\begin{figure}
\begin{center}
\includegraphics[width=0.5 \textwidth]{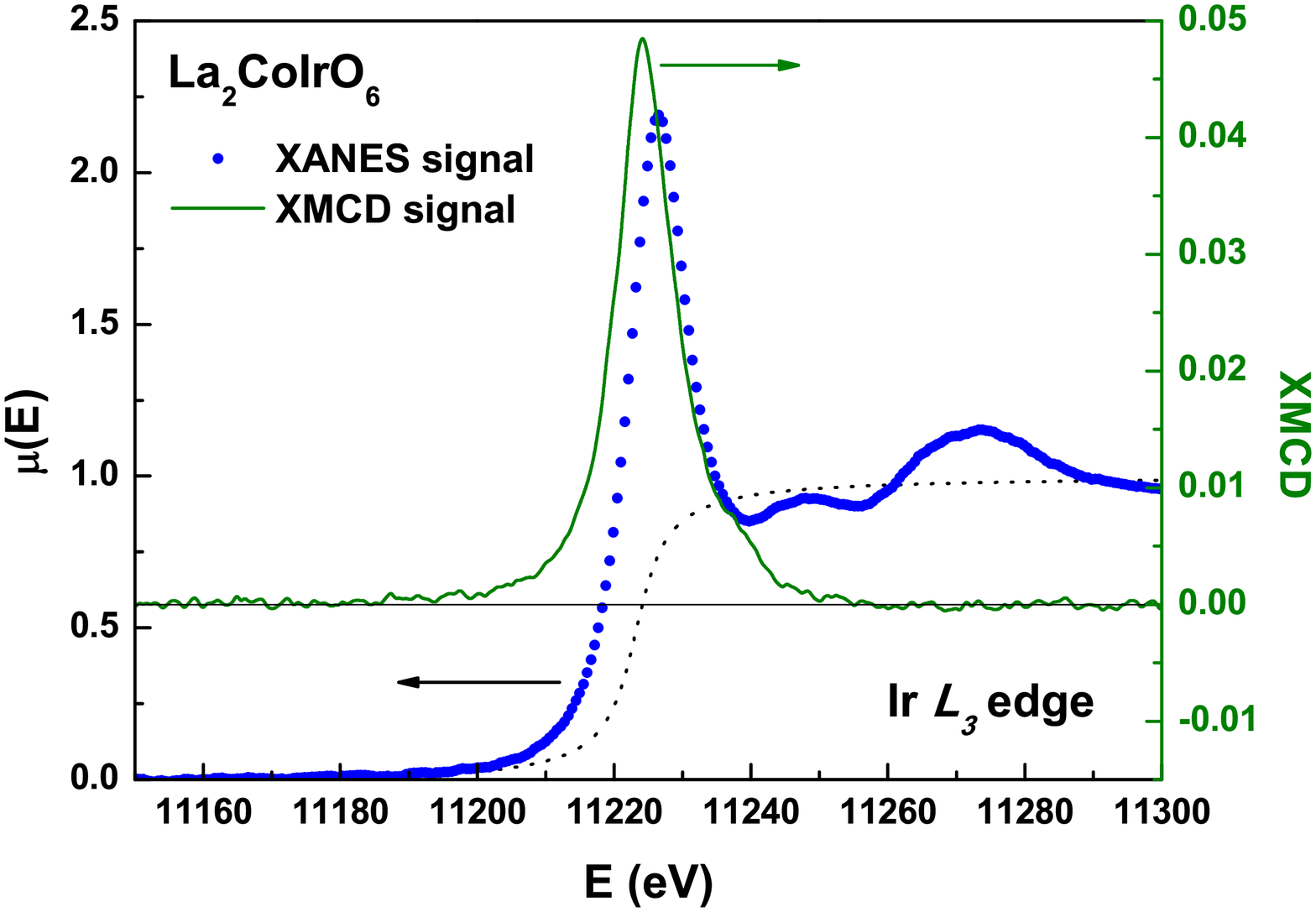}
\end{center}
\caption{XANES and XMCD spectrum at the Ir $L_3$ edge in La$_{2}$CoIrO$_{6}$.}
\label{FigS5}
\end{figure}

\begin{figure}
\begin{center}
\includegraphics[width=0.5 \textwidth]{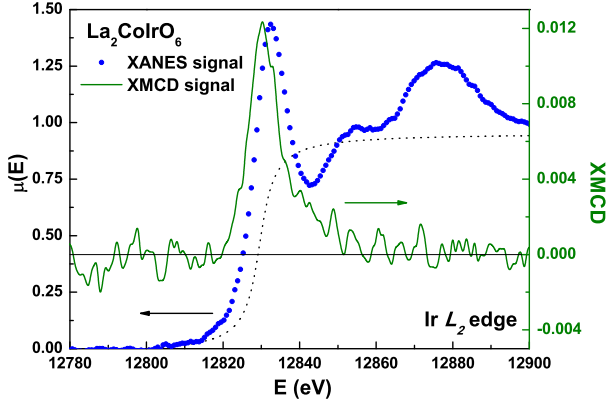}
\end{center}
\caption{XANES and XMCD spectrum at the Ir $L_2$ edge in La$_{2}$CoIrO$_{6}$.}
\label{FigS6}
\end{figure}

In Figs. \ref{FigS5} and \ref{FigS6} we present the XANES and XMCD spectra of La$_{2}$CoIrO$_{6}$ at the Ir $L_3$ and $L_2$ edges, respectively, at 60 K. By applying the well known sum rules which relates the integrated XAS and XMCD signals to polycrystalline samples (neglecting the magnetic dipole contribution \cite{s_xmcddipole}) we could extract the spin and orbital moments \cite{s_xmcdsum}.

For La$_{2}$CoIrO$_{6}$ we obtain for the Ir moments an orbital magnetic moment $\mu_{orb}=-0.14(1)$ $\mu_B$ and a spin magnetic moment $\mu_{spin}=-0.15(1)$ $\mu_B$, thus resulting in a total magnetic moment $\mu_{total}=-0.29(1)$ $\mu_B$ per Ir and $\mu_{orb}/\mu_{spin}=0.93$. These values are very close to the derived by Kolchinskaya \textit{et al.} \cite{s_Kolchinskaya}.

\begin{figure}
\begin{center}
\includegraphics[width=0.5 \textwidth]{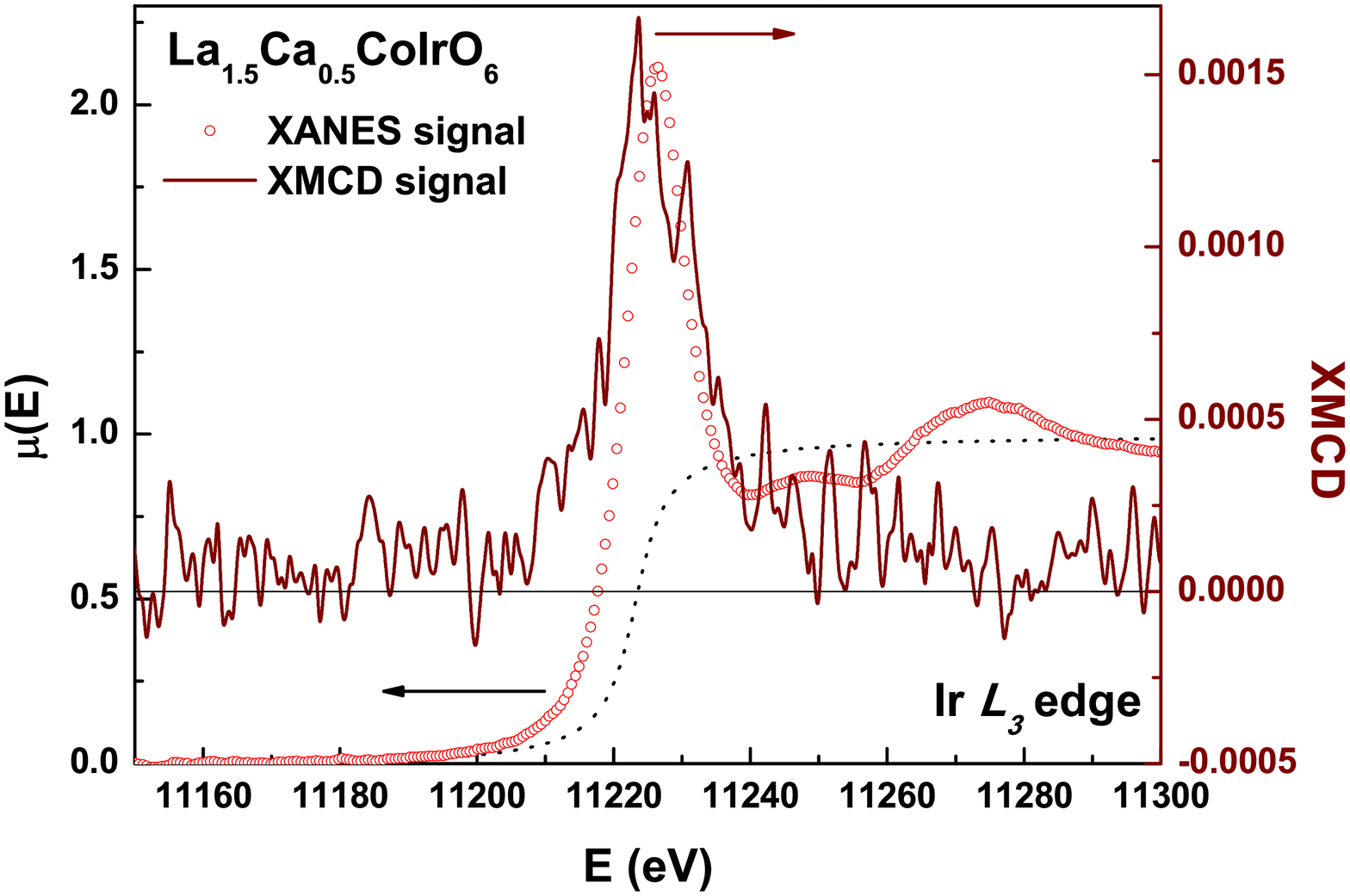}
\end{center}
\caption{XANES and XMCD spectrum at the Ir $L_3$ edge in La$_{1.5}$Ca$_{0.5}$CoIrO$_{6}$.}
\label{FigS7}
\end{figure}

Fig. \ref{FigS7} shows the XANES and XMCD spectra of La$_{1.5}$Ca$_{0.5}$CoIrO$_{6}$ at the Ir $L_3$ edge at 60 K. No XMCD signal was observed for $L_2$ edge, thus it was considered to be zero. Following the same procedure done for La$_{2}$CoIrO$_{6}$, we obtain for the Ir moments $\mu_{orb}=-0.009(1)$ $\mu_B$ and $\mu_{spin}=-0.013(1)$ $\mu_B$, thus $\mu_{total}=-0.022(1)$ $\mu_B$ per Ir and $\mu_{orb}/\mu_{spin}=0.69$. This very small value in respect to the expected moment for a $S=1/2$ confirms that the Ir$^{4+}$ ions are frustrated due to the AFM coupling of its Co$^{2+}$ neighbors.

\subsection*{Magnetometry}

\begin{figure}
\begin{center}
\includegraphics[width=0.4 \textwidth]{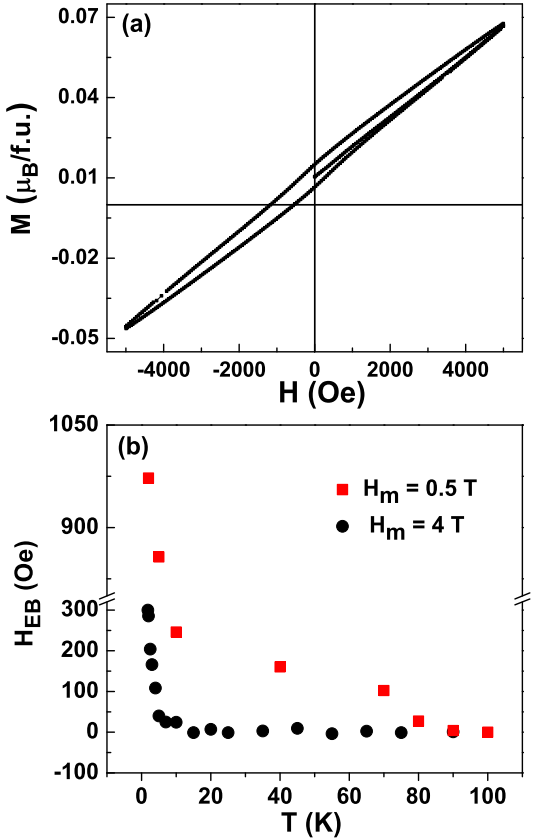}
\end{center}
\caption{(a) Zero field cooled (ZFC) $M(H)$ loop for $H_{m}=0.5$T at $T=5$ K. (b) $H_{eb}$ evolution with temperature for the ZFC magnetization as a function of applied field for the $H_{m}=0.5$T and $H_{m}=4$T $M(H)$ loops.}
\label{FigMinLoop}
\end{figure}

As discussed in the main text, a shift in the magnetization as a function of applied field can be observed in conventional FM/FIM materials if a minor $M(H)$ loop is measured, \textit{i.e.}, an $M(H)$ curve with the maximum applied field ($H_{m}$) not large enough to the system achieve the magnetic saturation.  These minor loops are usually highly asymmetric along both the field and magnetization axis. In order to further evidence that the EB observed for La$_{1.5}$Ca$_{0.5}$CoIrO$_{6}$ is not related to a minor loop effect, we measured minor loops with $H_{m}=0.5$ T at several temperatures. Fig. \ref{FigMinLoop}(a) shows the minor $M(H)$ loop measured at 5 K. As can be observed, the curve is highly asymmetric, exhibiting a large shift along the $M$ axis. This is completely different from the curve for $H_{m}=4$ T, which shows no appreciable asymmetry along the magnetization axis. Fig \ref{FigMinLoop}(b) shows the $T$-dependence of $H_{EB}=|H_{+}+H_{-}|/2$ for both the loops with $H_{m}=0.5$ T and $H_{m}=4$ T, for comparison. For the 4 T loops the $H_{EB}$ vanishes below $T_{f}$ while for the 0.5 T loops the shift persists above $T_{f}$ and only goes to zero on the paramagnetic state. If the $H_{EB}$ observed for the $H_{m}=4$ T loops were due to a minor loop effect, it should be expected to persist up to $\sim85$ K, in resemblance to the $H_{m}=0.5$ T curve. This remarkable difference between the curves indicate that the $H_{EB}$ observed for the $H_{m}=4$ T loop is an exchange biased phenomena which is intrinsically related to the RSG state.

Here is important to discuss the concept of zero magnetic field for the experimental apparatus used. In order to minimize the remanent magnetization on the magnet, the sample was always taken to the paramagnetic state from one measurement to another, and the magnetic field was sent to zero on the oscillating mode. But this does not warrant a precise zero field. To further confirm the results here described we have also heated up to room-$T$ and shut down the magnet, in order to ensure that there was no trapped flux on the magnet, and then performed ZFC measurements. For the checking $M(H)$ loops the sample exhibit the same EB effect, indicating that the unidirectional anisotropy and ZEB here described are intrinsic of the material and not due to some experimental artifact.

\subsection*{Electronic structure calculations}

The band structure calculations were performed using the WIEN2k software package \cite{s_Wien2k}. The FM and AFM cases were calculated with and without spin-orbit coupling on the Ir 5$d$ levels. The spin-orbit effect on 3$d$ levels is less important and was not included for the Co atoms. The exchange and correlation potential used was the PBEsol implementation of the GGA \cite{s_Perdew}. The wave function of the valence electrons was expanded using more than 193000 plane waves. The self consistent potential was obtained sampling 343 points in the Brillouin zone. The convergence criteria were set to 10$^{-5}$ eV on the total energy and 10$^{-3}$ $e^{-}$ on the electronic charges.

\begin{figure}
\begin{center}
\includegraphics[width=0.5 \textwidth]{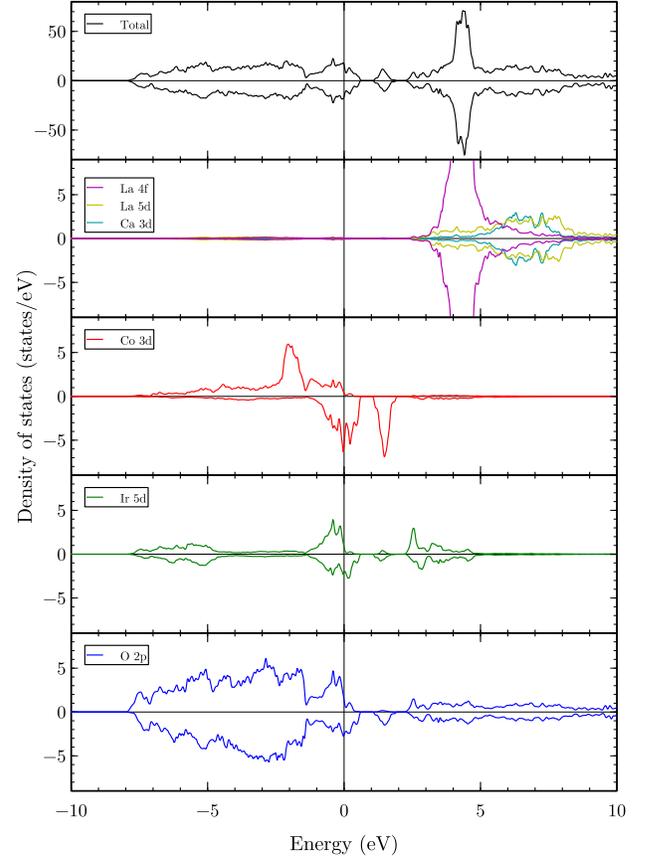}
\end{center}
\caption{Calculated total and partial density of states (DOS) for La$_{1.5}$Ca$_{0.5}$CoIrO$_{6}$ at AFM configuration.}
\label{FigDFT}
\end{figure}

\begin{table}
\centering
\caption{Relative energies and magnetic moments of La$_{1.5}$Ca$_{0.5}$CoIrO$_{6}$ obtained from band structure calculations for different magnetic structures. The energies are relative to the paramagnetic case.}
\label{T1}
\begin{tabular}{c|cccc}
\hline \hline
                  & $\Delta$E (eV) & $\mu_{Co}$ ($\mu_B$) & $\mu_{Ir}$ ($\mu_B$) & $\mu$ ($\mu_B$/f.u.) \\
\hline
Paramagnetic      & 0              & 0                    & 0                    & 0                    \\
\hline
FM        & -0.86           & 2.5                  & 0.6                  & 4.30                 \\
FM+SOC          & -6.46          & 2.4                  & 0.3                  & 3.48                 \\
\hline
AFM           & -0.93           & 2.4                  & 0.6                  & 0.07                \\
AFM+SOC   & -6.48          & 2.4                  & 0.3                  & 0.06                \\
\hline \hline
\end{tabular}
\end{table}

Table \ref{T1} gives the energies and magnetic moments of the La$_{1.5}$Ca$_{0.5}$CoIrO$_{6}$ compound in the PM, FM and AFM phases. The FM and AFM structures present lower total energies than the paramagnetic case. This shows that the magnetic interactions are very important in this material. Further, the FM and AFM energies decrease when spin-orbit coupling is included. Moreover, the magnetic moment of the Ir 5$d$ electrons shrink to half by the spin-orbit interaction. The AFM ordering is the most stable structure, although the FM arrangement is relatively close in energy. This indicates that the magnetic interactions present a large degree of frustration, and helps to explain the observed spin glass behavior in this material. Considering the $\sim$70\%/30\% of Co$^{2+}$/Co$^{3+}$ and the 9\% of ASD experimentally observed, and assuming the magnetic moments of the AFM and FM phases of Table \ref{T1}, the system's magnetization can be calculated as $M=M_{AFM}+M_{FM}=[(0.7+0.3\times0.91)m_{AFM}]+[(0.3\times0.09)m_{FM}]=0.15$ $\mu_{B}$/f.u, which is very close to the low-$T$ experimental values, $\sim0.13$ $\mu_{B}$/f.u..

Figure \ref{FigDFT} presents the total and partial densities of states for AFM La$_{1.5}$Ca$_{0.5}$CoIrO$_{6}$. All contributions are split into the majority and minority spin states and the zero energy corresponds to the Fermi level. Only the contributions of the spin up transition-metal ion and their ligands are shown (the spin down contribution are just complementary and do not add much to the discussion). The La 4$f$ states are extremely intense and were clipped to better visualize the other contributions. The Co 3$d$ and Ir 5$d$ states are split by the approximately $Oh$ crystal field into the $t_{2g}$ and $e_{g}$ structures. The O 2$p$ band appears strongly mixed with the metal states throughout the valence band, revealing a large covalent contribution to the bonding in this compound. This is especially true for the Ir 5$d$ states which show an even larger mixing due to the larger spatial extent of 5$d$ levels. The majority Co 3$d$ and Ir 5$d$ states present a minimum at the Fermi level, which resemble that observed in half-metallic magnetic materials. Finally, the La 4$f$, La 5$d$ and Ca 3$d$ states appear at much higher energies, exhibiting a mostly ionic contribution to the electronic structure.

The anti-site formation was also calculated, with a supercell with 80 inequivalent atomic sites. This gives an anti-site concentration of about 12$\%$, in line with the estimated 9$\%$ concentration. The difference in total energy between the disordered and perfect lattice was about 1.1 eV. This is similar to the 0.94 eV result obtained for anti-site formation in the related LnBaCoO$_{5.5}$ double-perovskite \cite{s_Seymour}.


\end{document}